\documentclass[prd,two column]{revtex4}

\usepackage{graphicx}
\usepackage{amsmath}
\usepackage{comment}
\usepackage{slashed}
\usepackage{CJKutf8}
\usepackage{url}
\usepackage{bm}
\usepackage{amssymb}

\usepackage{tikz}
\newcommand*{\circled}[1]{\lower.7ex\hbox{\tikz\draw (0pt, 0pt)%
		circle (.5em) node {\makebox[1em][c]{\small #1}};}}

\usepackage{tikz}
\usetikzlibrary{arrows,shapes}
\usetikzlibrary{trees}
\usetikzlibrary{matrix,arrows}
\usetikzlibrary{positioning}				
\usetikzlibrary{calc,through}				
\usetikzlibrary{decorations.pathreplacing}  

\usepackage{pgffor}							
\usetikzlibrary{decorations.pathmorphing}	
\usetikzlibrary{decorations.markings}

\tikzstyle{block} = [draw, rectangle, minimum height=3em, minimum width=6em]
\usepackage{hyperref}
\usepackage{amsmath}
\usepackage{ulem}

\usetikzlibrary{decorations.pathmorphing}	
\usetikzlibrary{decorations.markings}
\tikzset{
	doublearrow/.style={draw=black,double =black,double distance=2.0pt,
		postaction={decorate},decoration={markings,mark=at position 1.0 with {\arrow[draw=black,line width=1.5pt]{>}}}},
	heavyquark/.style={draw=black,double =white,double distance=2.0pt,
		postaction={decorate},
		decoration={markings,mark=at position .55 with {\arrow[draw=black,line width=1.5pt]{>}}}},
	heavyquarkbar/.style={draw=black,double =white,double distance=2.0pt,postaction={decorate},
		decoration={markings,mark=at position .55 with {\arrow[draw=black,line width=1.5pt]{<}}}},
	relevant_quark/.style={draw=black,double =white,double distance=2.0pt, postaction={decorate}, decoration={markings,mark=at position .55 with {\arrow[draw=black,line width=2.5pt]{>}}}},
	irrelevant_quark/.style={draw=black,double =white,double distance=2.0pt,postaction={decorate}, decoration={markings,mark=at position .485 with {\arrow[draw=black,line width=2.5pt]{>}},mark=at position .665 with {\arrow[draw=black,line width=2.5pt]{>}}}},
	solidline/.style={draw=black, postaction={decorate}},
	thickline/.style={draw=black, line width=4.0pt, postaction={decorate}},
	thickarrow/.style={draw=black, line width=4.0pt,postaction={decorate},decoration={markings,mark=at position .63 with {\arrow[draw=red,line width=2.5pt]{>}}}},
	thick_2_arrow/.style={draw=black, line width=4.0pt,postaction={decorate},decoration={markings,mark=at position .73 with {\arrow[draw=red,line width=2.5pt]{>>}}}},
	doubleline/.style={draw=black,double =white,double distance=2.0pt,postaction={decorate}},
	vector/.style={decorate, decoration={snake,amplitude=1.0pt, segment length=4pt}, draw},
	provector/.style={decorate, decoration={snake,amplitude=2.5pt}, draw},
	antivector/.style={decorate, decoration={snake,amplitude=-2.5pt}, draw},
	fermion/.style={draw=black, postaction={decorate},
		decoration={markings,mark=at position .55 with {\arrow[draw=black]{>}}}},
	fermionbar/.style={draw=black, postaction={decorate},
		decoration={markings,mark=at position .55 with {\arrow[draw=black]{<}}}},
	fermionnoarrow/.style={draw=black},
	gluon/.style={decorate, draw=black,
		decoration={coil,amplitude=4pt, segment length=5pt}},
	scalar/.style={dashed,draw=black, postaction={decorate},
		decoration={markings,mark=at position .55 with {\arrow[draw=black]{>}}}},
	scalarbar/.style={dashed,draw=black, postaction={decorate},
		decoration={markings,mark=at position .55 with {\arrow[draw=black]{<}}}},
	scalarnoarrow/.style={dashed,draw=black},
	electron/.style={draw=black, postaction={decorate},
		decoration={markings,mark=at position .55 with {\arrow[draw=black]{>}}}},
	bigvector/.style={decorate, decoration={snake,amplitude=4pt}, draw},
	arrow/.style={draw=black, postaction={decorate},
		decoration={markings,mark=at position 1. with {\arrow[draw=black]{>}}}},
}
\tikzstyle{block} = [draw, rectangle, 
minimum height=3em, minimum width=6em]

\begin{document}
\title{Treating divergence in quark matter by using energy projectors}
\date{\today}
	
\author{Guojun Huang and Pengfei Zhuang}
\address{Physics Department, Tsinghua University, Beijing 100084, China}
\begin{abstract}
We calculate gluon self-energy using quark energy projectors in a general quark-gluon plasma. By separating the quark field into a positive- and a negative-energy mode, the quark loop constructed with the same mode is always convergent, and the divergence appears only in the mixed loop with different modes and is medium independent. After removing the divergence in vacuum, we obtain the one-loop gluon self-energy at finite temperature, chemical potential and quarks mass without approximation. With the method of quark loop resummation, we calculate non-perturbatively the gluon Debye mass and thermodynamic potential. In the limit of small gluon momentum in comparison with temperature, chemical potential and quark mass, our calculation comes back to the known HTL/HDL results in literature. 
\end{abstract}
\maketitle

\section{Introduction}
The properties of QCD matter at finite temperature and chemical potential, especially the deconfinement and chiral symmetry phase transitions~\cite{Alford:1997zt,Shuryak:1980tp,Stephanov:1998dy,Alford:2007xm,Rajagopal:2000wf,deForcrand:2002hgr,Karsch:2001cy,CasalderreySolana:2011us,Cassing:1999es,Allton:2002zi,Fukushima:2010bq,Berges:1998rc,Halasz:1998qr,DElia:2002tig,Karsch:2000kv,Son:2000xc,Rajagopal:1992qz,Stephanov:2004wx,Fukushima:2008wg,Blaizot:2001nr,deForcrand:2010ys,Tawfik:2021eeb,BraunMunzinger:2009zz,Bazavov:2018mes,Schmidt:2017bjt,Fu:2010pv} and their realization in high energy nuclear collisions~\cite{Albacete:2014fwa,Iancu:2012xa,Aoki:2009sc,Harris:1996zx,Jacobs:2004qv,Zschiesche:2002zr,Rapp:2008qc,Yee:2013cya,Mohanty:2013yca,Redlich:2012xf,She:2017icp,Luo:2020pef,Gastineau:2004ad} and compact stars~\cite{Akmal:1998cf,Buballa:2003qv,Lattimer:2000nx,Wiringa:1988tp,Abbott:2018exr,Li:1998bw,Friedman:1981qw,Douchin:2001sv,Glendenning:1992vb,Weber:2004kj,Baym:1971ax,Oertel:2016bki,Carlson:2014vla,Burgio:2021vgk,Friman:2014cua} are widely studied for decades. As a often used method in theoretical calculations, one introduces energy projectors to divide the quark field into a positive- and a negative-energy modes, and therefore any Feynman diagram is separated into two groups: the pure fraction constructed by the same modes and the mixed fraction by the two modes~\cite{Hong:1998tn,Hong:2004qf,Schafer:2003jn,Schafer:2004yx,Schafer:2004zf,Nardulli:2002ma,Huang:2001yw}. The energy-projector method is widely used in the study of color superconductivity at extremely high baryon density~\cite{Nardulli:2002ma,Anglani:2011cw,Hong:2004qf,Casalbuoni:2002st,Ruggieri:2003nu}. In non-relativistic limit with heavy quark mass (NRQCD), the positive- and negative-energy modes become respectively the relevant and irrelevant modes, and by integrating out the irrelevant mode, the relevant mode becomes the dominant one and controls the behavior of heavy quark systems~\cite{brambilla1,isgur,brambilla2,luke,braaten,goncalves,weinberg,richard,manohar,NRQED_to_1overm4,dye,gerlach,Bodwin:1994jh,Eichten:1989zv,Isgur:1989vq,Isgur:1990yhj,Grinstein:1990mj,Georgi:1990um}.     

One-loop diagram plays a crucial role in non-perturbative calculations of QCD, like the approaches of hard thermal loop resummation (HTL)~\cite{Andersen:1999fw,Haque:2014rua,Andersen:1999sf,Andersen:1999va,Peshier:2000hx,Peshier:1998dy} and hard dense loop resummation (HDL)~\cite{Jiang:2010jm}. A key problem in these calculations is how to treat the divergence at one-loop level. In some limits like extremely high temperature or high baryon density, the divergence is properly removed in HTL and HDL~\cite{TFT_Bellac1996,Kapusta:2006pm,Blaizot:2000fc}.{\color{red}} We focus in this paper on the divergence problem in the calculation of in-medium one-loop gluon self-energy at finite temperature, chemical potential and quark mass, using the quark energy-projector method. We will see that, the divergence appears only in the mixed loop and is medium independent. Therefore, it does not change the thermodynamic properties relative to the vacuum and can be directly removed.

The paper is organized as follows. We rewrite the quark sector of the QCD Lagrangian density in terms of energy projectors in Nambu-Gorkov space in Section \ref{sec2}, and then calculate the gluon self-energy at one-loop level by separating the quark loop into a pure loop without divergence and a mixed loop with vacuum divergence in Section \ref{sec3}. After taking the often used loop resummation, we calculate in Section \ref{sec4} the thermodynamic properties of the quark matter, like the gluon Debye mass and gluon thermodynamic potential, and compare our calculations under some extreme conditions with the known results in literature. We finally summarize in Section \ref{sec5}. 

\section{Energy-projector method}
\label{sec2}
We first consider particle-antiparticle symmetry of strong interaction. By taking the charge conjugation operator $C$ which changes quark fields $\psi(x)$ and $\bar\psi(x)$ to $\psi_C(x)=C\bar\psi^T(x)$ and $\bar\psi_C(x)=\psi^T(x)C$, one introduces the Nambu-Gorkov space~\cite{Rischke:2000qz,Rischke:2003mt}
\begin{equation}
\begin{aligned}
\Psi = \left(\begin{matrix}
\psi\\ \psi_C
\end{matrix}\right),\ \ \ \bar\Psi = (\bar\psi,\bar\psi_C).
\end{aligned}
\end{equation}
To make the fields in momentum space dimensionless, the normalization factors in the Fourier transformation from coordinate space to momentum space for the quark field $\Psi$ and gluon field $A_\mu^a$ are chosen as
\begin{eqnarray}
&& \Psi(x) = {1\over\sqrt V}\sum_k e^{-ik\cdot x}\Psi(k),\nonumber\\
&& A_\mu^a(x) = {1\over\sqrt{TV}}\sum_q e^{-iq\cdot x}A_\mu^a(q),
\end{eqnarray}
where $V$ and $T$ are the volume and temperature of the thermal system, $k_0=-i(2n_k+1)\pi T$ and $q_0 =-i2n_q\pi T$ with $n_k,n_q=0,\pm 1, \pm 2,\cdots$ appeared in $k\cdot x=k_0t -{\bf k}\cdot{\bf x}$ and $q\cdot x=q_0t-{\bf q}\cdot{\bf x}$ are the quark and gluon frequencies in the imaginary time formalism of finite temperature field theory, and the summations $\sum_k$ and $\sum_q$ mean the frequency summation and momentum integration. 

The quark sector of QCD Lagrangian density in coordinate space
\begin{equation}
{\cal L}=\bar\psi\left(i\gamma^\mu\partial_\mu+g\gamma^\mu A_\mu^aT_a+\mu_f\gamma_0-m_f\right)\psi
\end{equation}
with quark mass $m_f$, quark chemical potential $\mu_f$, Gell-Mann matrices $T_a\ (a=0,1,\cdots,8)$ and quark-gluon coupling constant $g$ can be expressed as 
\begin{equation}
\label{l1}
{\cal L} = {1\over 2}\sum_p \bar\Psi(k)G^{-1}(k,p)\Psi(p)
\end{equation}
in momentum space and Nambu-Gorkov space, where the full quark propagator 
\begin{equation}
G^{-1} = G_0^{-1}+g{\cal A}
\end{equation}
contains the free propagator 
\begin{eqnarray}
&& \begin{aligned} G_0^{-1}(k,p) = {1\over T}\left(\begin{matrix}[G_0^+]^{-1}(k)&0\\0&[G_0^-]^{-1}(k)\end{matrix}\right)\delta(k-p) \end{aligned},\nonumber\\
&& [G_0^\pm]^{-1}(k) = \gamma^\mu k_\mu\pm\mu_f\gamma_0-m_f
\end{eqnarray}
and the modified gauge field 
\begin{eqnarray}
\label{a}
&& {\cal A}(k,p) = {1\over\sqrt{VT^3}}\Gamma^\mu_a A_\mu^a(k-p),\nonumber\\
&& \Gamma^\mu_a = \gamma^\mu\left(\begin{matrix} T_a &0\\ 0&-T_a^T \end{matrix}\right). 
\end{eqnarray}

We further separate the quark fields into two parts with positive and negative energy. The energy projectors onto states of positive and negative energy for free massive quarks are defined as~\cite{Pisarski:1999av,Pisarski:1999tv,Huang:2001yw,Reuter:2004kk,Rho:2000ww}
\begin{equation}
\Lambda_\pm(\tilde k) = {1\over 2\epsilon_k}\left[\epsilon_k\pm\gamma_0({\bm\gamma}\cdot{\bf k}+m_f)\right]
\end{equation}
with the quark energy $\epsilon_k=\sqrt{m_f^2+{\bf k}^2}$. Note that, $\tilde k=(\epsilon_k,{\bf k})$ is a on-shell four momentum which is different from the general four momentum $k=(k_0,{\bf k})$. Taking into account the orthogonal and complete properties,
\begin{equation}
\Lambda_+\Lambda_-=\Lambda_-\Lambda_+=0,\ \ \ \Lambda_+ + \Lambda_- = 1
\end{equation}  
and the relations 
\begin{equation}
\Lambda_\pm^\dag=\Lambda_\pm,\ \ \ \ \Lambda_\pm^2=\Lambda_\pm, 
\end{equation}
it is easy to check that the states 
\begin{equation}
\Psi_\pm(k) = \Lambda_\pm (\tilde k)\Psi(k)
\end{equation}
satisfy the Dirac equation 
\begin{equation}
H\Psi_\pm (k) =\pm \epsilon_k\Psi_\pm (k)
\end{equation}
with the free Hamiltonian $H=\gamma_0({\bm \gamma}\cdot{\bf k}+m_f)$. This is the reason why we call $\Psi_\pm$ the positive- and negative-energy states. Using the energy projectors, the Lagrangian density (\ref{l1}) can be rewritten as  
\begin{equation}
\label{l2}
{\cal L}  ={1\over 2}\sum_{m,n=\pm}\sum_p\bar\Psi_m(k)G_{mn}^{-1}(k,p)\Psi_n(p)
\end{equation}
with the matrix elements of the full quark propagate in energy space, 
\begin{eqnarray}
\label{ga}
&& G^{-1}_{mn}(k,p) = [G_0^{-1}]_{mn}(k,p)+g{\cal A}_{mn}(k,p),\nonumber\\
&& [G_0^{-1}]_{mn}(k,p) = \gamma_0\Lambda_m(\tilde k)\gamma_0 G_0^{-1}(k,p)\Lambda_n(\tilde p),\nonumber\\
&& {\cal A}_{mn}(k,p) = \gamma_0\Lambda_m(\tilde k)\gamma_0{\cal A}(k,p)\Lambda_n(\tilde p). 
\end{eqnarray}
The matrix elements of the free propagate can be explicitly expressed as 
\begin{eqnarray}
\label{g0}
&& [G_0^{-1}]_{++}(k,p) = {1\over T}\left(\begin{matrix} 	k_0+\mu_f-\epsilon_k&0\\ 0&k_0-\mu_f-\epsilon_k \end{matrix}\right)\nonumber\\
&&\ \ \ \ \ \ \ \ \ \ \ \ \ \ \ \ \ \ \ \times\gamma_0\Lambda_+(\tilde k)\delta(k-p),\nonumber\\
&& [G_0^{-1}]_{--}(k,p) = {1\over T}\left(\begin{matrix} 	k_0+\mu_f+\epsilon_k&0\\ 0&k_0-\mu_f+\epsilon_k \end{matrix}\right)\nonumber\\
&&\ \ \ \ \ \ \ \ \ \ \ \ \ \ \ \ \ \ \ \times\gamma_0\Lambda_-(\tilde k)\delta(k-p),\nonumber\\
&& [G_0^{-1}]_{+-}(k,p) = [G_0^{-1}]_{-+}(k,p)=0.
\end{eqnarray}
While the free propagator is diagonal in energy space, the gauge field is with off-diagonal elements which lead to the coupling between the positive- and negative-energy fields $\Psi_+$ and $\Psi_-$. 

In non-relativistic QCD theory (NRQCD) for heavy quarks, the particles with negative energy are irrelevant modes and can be integrated out~\cite{Kilian:1993mw}. By redefining a new negative-energy field instead of $\Psi_-$ to remove the cross terms between $\Psi_+$ and $\Psi_-$, and then integrating out the new field, the Lagrangian density contains only the positive-energy field $\Psi_+$. However, for light flavors both the positive- and negative-energy fields $\Psi_+$ and $\Psi_-$ are relevant modes and should be treated equally importantly.   

We now extract, from the Lagrangian density (\ref{l2}), the Feynman rules for the positive- and negative-energy quark fields in Nambu-Gorkov space, which will be used in the calculation of gluon self-energy later. Considering the diagonal property (\ref{g0}) of the free propagator $G_0^{-1}$ in energy space, the Lagrangian density becomes
\begin{eqnarray}
\label{l3}
{\cal L} &=& {1\over 2}\sum_p\Big[\sum_{m=\pm}\bar\Psi_m(k)[G_0^{-1}]_{mm}(k,p)\Psi_m(p)\nonumber\\
&& + g\sum_{m,n=\pm}\bar\Psi_m(k){\cal A}_{mn}(k,p)\Psi_n(p)\Big].
\end{eqnarray}
By taking into account the relations (\ref{a}), (\ref{ga}) and (\ref{g0}) for the quark and gauge fields and the projection properties 
\begin{equation}
\Lambda_\pm \Psi_\pm = \Psi_\pm,\ \ \ \ \Lambda_\mp \Psi_\pm = 0, 
\end{equation}
the two free propagators for the positive- and negative-energy quarks can be expressed as
\begin{equation}
\label{pro}
{\mathbb G}_0^\pm(k) = T\left(\begin{matrix} {1\over k_0+\mu_f\mp \epsilon_k}&0\\ 0&{1\over k_0-\mu_f\mp \epsilon_k} \end{matrix}\right)
\end{equation}
which satisfy the completeness condition
\begin{equation}
\left[\sum_n{\mathbb G}_0^n\Lambda_n\right]\gamma_0=G_0,
\end{equation} 
and the four kinds of coupling vertexes among gluon and positive- and negative-energy quarks can be generally represented as  
\begin{equation}
\label{ver}
{\mathbb V}_{mn}^{\mu,a}(\tilde k,\tilde p) = {g\over \sqrt{VT^3}}\Lambda_m(\tilde k)\gamma^0\Gamma^\mu_a\Lambda_n(\tilde p).
\end{equation}
   
\section{One-loop gluon self-energy }
\label{sec3}
\begin{figure}
\centering
\includegraphics[height=1cm,width=8cm]{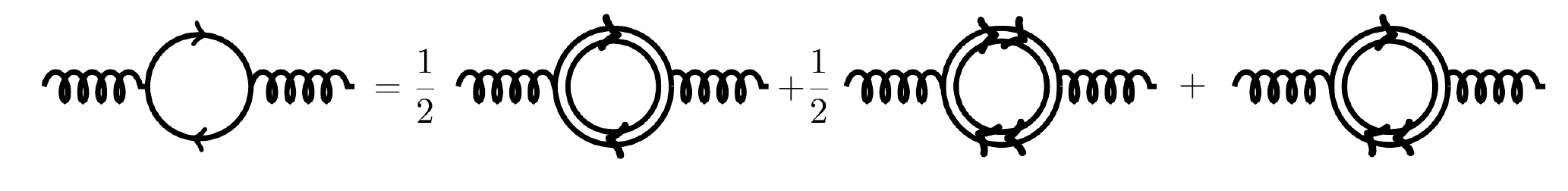}
\caption{The gluon self-energy at one quark loop level. The double lines indicate quark modes with positive (one arrow) and negative (double arrows) energies. }
\label{fig1}
\end{figure}

We calculate in this section the gluon self-energy, using the quark propagators (\ref{pro}) and coupling vertexes (\ref{ver}). Since a quark loop can be constructed by two positive-energy quarks, two negative-energy quarks and one positive- and one negative-energy quarks, the quark contribution $\Pi^{\mu\nu,ab}_Q(q)$ to the gluon self-energy $\Pi^{\mu\nu,ab}(q)$ at one loop level contains three parts, see Figure \ref{fig1},
\begin{eqnarray}
\Pi^{\mu\nu,ab}_Q &=& \Pi^{\mu\nu,ab}_{++} + \Pi^{\mu\nu,ab}_{--} + 2\Pi^{\mu\nu,ab}_{+-},\\
\Pi^{\mu\nu,ab}_{mn}(q) &=& -{1\over2}{(-1)^2\over 2!}(2-1)!\sum_k\text{Tr}\Big[{\mathbb G}_0^m(k_+)\nonumber\\
&&\times{\mathbb V}^{\mu,a}_{mn}(\tilde k_+,\tilde k_-){\mathbb G}_0^n(k_-){\mathbb V}^{\nu,b}_{nm}(\tilde k_-,\tilde k_+)\Big]\nonumber
\end{eqnarray}
with two quark momenta $k_+^\mu = (k_0,{\bf k}+ {\bf q}/2)$, $k_-^\mu = (k_0-q_0,{\bf k}- {\bf q}/2)$, on-shell momenta $\tilde k_\pm^{\mu} = (\epsilon_\pm,{\bf k}\pm {\bf q}/2)$ and energies $\epsilon_\pm = \sqrt{m_f^2+({\bf k}\pm{\bf q}/2)^2}$. The first coefficient $1/2$ in $\Pi^{\mu\nu,ab}_{mn}(q)$ is from the normalization in Nambu-Gorkov space, and the following coefficient ${(-1)^n}(n-1)!/n!$ $(n=2)$ is from the topological number of the Feynman diagram. We have used here the symmetry $\Pi^{\mu\nu,ab}_{+-}=\Pi^{\mu\nu,ab}_{-+}$ in energy space. 

Taking into account the projection properties
\begin{equation}
{\mathbb G}_0^n\Lambda_n=\Lambda_n{\mathbb G}_0^n
\end{equation} 
and $\Lambda_n^2=\Lambda_n$, there is
\begin{eqnarray}
\Pi^{\mu\nu,ab}_{mn}(q) &=&-{g^2\over 4VT^3}\sum_k\text{Tr}\Big[{\mathbb G}_0^m(k_+)\Lambda_m(\tilde k_+)\gamma^0\Gamma_a^\mu\nonumber\\
&&\times{\mathbb G}_0^n(k_-)\Lambda_n(\tilde k_-)\gamma^0\Gamma_b^\nu\Big].
\end{eqnarray}

\subsection{$\Pi_{--}^{\mu\nu,ab}(q)$}
We now take $\Pi^{\mu\nu,ab}_{--}$ as an example to show the calculation and simplification of the gluon self-energy. After considering the trace in color space $\text{Tr}\left(T_aT_b\right)=\text{Tr}\left(T_a^TT_b^T\right)=\delta_{ab}/2$, the self-energy in color and spin spaces at one-loop level can be factorized as two parts,
\begin{eqnarray}
\label{cs}
&& \Pi^{\mu\nu,ab}_{--} = -{1\over T^2}{\delta_{ab}\over 2}\Pi^{\mu\nu}_{--},\\
&& \Pi_{--}^{\mu\nu} =  {g^2T\over 4V}\sum_{k,s=\pm}{\text{Tr}\left[\Lambda_-(\tilde k_+)\gamma^0\gamma^\mu\Lambda_-(\tilde k_-)\gamma^0\gamma^\nu\right]\over \left(k_+^0+s\mu_f+\epsilon_+\right)\left(k_-^0+s\mu_f+\epsilon_-\right)},\nonumber
\end{eqnarray}
where the summation $\sum_{s=\pm}$ is over the quark and anti-quark in Nambu-Gorkov space. By considering the trace in spin space, 
\begin{eqnarray}
\label{trace1}
&& \text{Tr}\left[\Lambda_-(\tilde k_+)\gamma^0\gamma^\mu\Lambda_-(\tilde k_-)\gamma^0\gamma^\nu\right]\nonumber\\
&=& {1\over \epsilon_+\epsilon_-}\left[g^{\mu\nu}\left(m_f^2-\bar k_+^\sigma \bar k^-_\sigma\right)+\bar k_+^\mu \bar k_-^\nu+\bar k_+^\nu \bar k_-^\mu\right],
\end{eqnarray}
where the two new on-shell quark momenta $\bar k_\pm$ are defined as $\bar k_\pm = (-\epsilon_\pm, {\bf k}\pm{\bf q}/2)$, and summarizing the quark frequencies,
\begin{eqnarray}
\label{f}
&& \sum_{k_0,s=\pm}{1\over\left(k_+^0+s\mu_f+\epsilon_+\right)\left(k_-^0+s\mu_f+\epsilon_-\right)}\nonumber\\
&=& {\sum_{s=\pm}\left[f_F(\epsilon_++s\mu_f)-f_F(\epsilon_-+s\mu_f)\right]\over T\left(q_0+\epsilon_+-\epsilon_-\right)}\nonumber\\
&\equiv& {\epsilon_-\epsilon_+\over 2T}F(q_0,\epsilon_+,\epsilon_-)
\end{eqnarray}
with the Fermi-Dirac distribution $f_F(x)=1/(e^{x/T}+1)$, where we have used the relation $\tanh(x+in_q\pi)=\tanh(x)$, the gluon self-energy in spin space $\Pi^{\mu\nu}_{--}(q)$ can be further separated into two parts,
\begin{equation}
\Pi^{\mu\nu}_{--} = g^{\mu\nu}\overline\Pi_{--}+ \overline\Pi_{--}^{\mu\nu}
\end{equation}
with the scalar function $\overline\Pi_{--}(q)$ and tensor function $\overline\Pi_{--}^{\mu\nu}(q)$, 
\begin{eqnarray}
\label{Pi}
\overline\Pi_{--} &=& {g^2\over 8}\int{d^3{\bf k}\over (2\pi)^3}\left(m_f^2-\bar k_+^\sigma \bar k^-_\sigma\right)F(q_0,\epsilon_+,\epsilon_-),\\
\overline\Pi_{--}^{\mu\nu} &=& {g^2\over 8}\int{d^3{\bf k}\over (2\pi)^3} \left(\bar k_+^\mu \bar k_-^\nu +\bar k_+^\nu \bar k_-^\mu\right)F(q_0,\epsilon_+,\epsilon_-),\nonumber
\end{eqnarray}
where we have taken the continuous integration $\int d^3{\bf k}/(2\pi)^3$ over the quark three-momentum ${\bf k}$, instead of the discrete summation $\sum_{\bf k}/V$. Note that, the function $F(q_0, \epsilon_+, \epsilon_-)$ defined in (\ref{f}) contains only the Fermi-Dirac distribution, and therefore any ultraviolet divergence of the integral will be suppressed by the exponential function $e^{-\epsilon_\pm/T}$ in the distribution and the integral is always convergent.   

A key question for the calculation of the gluon self-energy is the divergence analysis. If there exists any infrared or ultraviolet divergence, a renormalization procedure is required. To see this clearly, we further simplify the three-momentum integration in equation (\ref{Pi}). We take the component $k_z$ of the quark momentum ${\bf k}$ along the gluon momentum ${\bf q}$,  
\begin{equation}
{\bf k} = |{\bf k}|\cos\theta\hat{\bf q}+|{\bf k}|\sin\theta\cos\phi\hat{\bf k}_x+|{\bf k}|\sin\theta\sin\phi\hat{\bf k}_y, 
\end{equation}
where ${\bf k}_x$ and ${\bf k}_y$ are perpendicular to the gluon momentum ${\bf q}$. Taking into account the symmetry in the transverse plane, $\Pi_{--}^{\mu\nu}(q)$ is independent of the choice of the directions $\hat{\bf k}_x$ and $\hat{\bf k}_y$. The integration over the azimuth angle $\phi$ is easy. Considering that the quark energies $\epsilon_\pm=\sqrt{m_f^2+|{\bf k}|^2+|{\bf q}|^2/4\pm |{\bf k}||{\bf q}|\cos\theta}$ and in turn the function $F(q_0,\epsilon_+,\epsilon_-)$ are independent of the angle $\phi$, the scalar and tensor functions $\overline\Pi_{--}(q)$ and $\overline\Pi_{--}^{\mu\nu}(q)$ can be easily written as 
\begin{eqnarray}
\overline\Pi_{--}  &=& {g^2\over 32\pi^2}\int d|{\bf k}| d\cos\theta |{\bf k}|^2\left(m_f^2-\bar k_+^\mu \bar k^-_\mu\right)\\
&&\times F(q_0,\epsilon_+,\epsilon_-),\nonumber\\
\overline\Pi_{--}^{\mu\nu}  &=& {g^2\over 32\pi^2}\int d|{\bf k}| d\cos\theta |{\bf k}|^2H^{\mu\nu}(q,k)F(q_0,\epsilon_+,\epsilon_-),\nonumber\\
H^{00} &=& 2\epsilon_+\epsilon_-,\nonumber\\
H^{0i} &=& H^{i0}= -\sum_{n=\pm}\epsilon_n\left(|{\bf k}|\cos\theta-n{|{\bf q}|\over 2}\right)\hat q^i,\nonumber\\
H^{ij} &=& \left(2 {\bf k}^2 \cos ^2\theta-{1\over 2}|{\bf q}|^2\right)\hat q^i \hat q^j+{\bf k}^2 \sin ^2\theta(\delta^{ij}-\hat q^i\hat q^j)\nonumber
\end{eqnarray}
with the function $F(q_0,\epsilon_+,\epsilon_-)$ defined in (\ref{f}).

Then we do variable substitution from $(|{\bf k}|,\cos\theta)$ to $(\epsilon_-,\epsilon_+)$. From the relations 
\begin{eqnarray}
&& {\bf k}^2 = (\epsilon_+^2+\epsilon_-^2)/2-m_f^2-{\bf q}^2/4,\nonumber\\
&& \cos\theta ={\epsilon_+^2-\epsilon_-^2 \over 2|{\bf q}||{\bf k}|}
\end{eqnarray}
and the corresponding Jacobian determinant
\begin{equation}
\left|{\partial(|{\bf k}|,\cos\theta)\over\partial(\epsilon_-,\epsilon_+)}\right| = {\epsilon_-\epsilon_+\over|{\bf q}||{\bf k}|},
\end{equation}
we finally obtain the scalar and tensor parts of the one-loop gluon self-energy in terms of the integration over $\epsilon_-$ and $\epsilon_+$,
\begin{eqnarray}
\label{int}
\overline\Pi_{--} &=& {g^2\over 32\pi^2}\int_R d\epsilon_-d\epsilon_+{\epsilon_-\epsilon_+\over |{\bf q}|}{(\epsilon_--\epsilon_+)^2-{\bf q}^2\over 2}\nonumber\\
&&\times F_S(q_0,\epsilon_+,\epsilon_-),\nonumber\\
\overline\Pi_{--}^{\mu\nu} &=& {g^2\over 32\pi^2}\int_R d\epsilon_-d\epsilon_+{\epsilon_-\epsilon_+\over |{\bf q}|}H^{\mu\nu}(q,\epsilon_+,\epsilon_-)\nonumber\\
&& \times F_S(q_0,\epsilon_+,\epsilon_-),\nonumber\\
H^{00} &=& 2\epsilon_+\epsilon_-,\nonumber\\
H^{0i} &=& H^{i0} ={(\epsilon_++\epsilon_-)^2-{\bf q}^2\over 2|{\bf q}|}q_0\hat q^i,\nonumber\\
H^{ij} &=& {1\over 2}\left({(\epsilon_-^2-\epsilon_+^2)^2\over {\bf q}^2}-{\bf q}^2\right)\hat q^i\hat q^j\nonumber\\
&& +\left[{1\over 2}(\epsilon_-^2+\epsilon_+^2-2m_f^2)-{1\over 4}\left({(\epsilon_-^2-\epsilon_+^2)^2\over {\bf q}^2}+{\bf q}^2\right)\right]\nonumber\\
&&\times(\delta^{ij}-\hat q^i\hat q^j),
\end{eqnarray}
where the integration region $R$ is controlled by the requirement that the three momenta ${\bf q}, {\bf k}+{\bf q}/2$ and ${\bf k}-{\bf q}/2$ should form a triangle, which leads to the constraints on the quark energies,
\begin{eqnarray}
&& \sqrt{\epsilon_-^2-m_f^2}+\sqrt{\epsilon_+^2-m_f^2}\ge |{\bf q}|,\nonumber\\
&& \left|\sqrt{\epsilon_-^2-m_f^2}-\sqrt{\epsilon_+^2-m_f^2}\right|\le |{\bf q}|,
\end{eqnarray}
and the function $F_S$ defined as 
\begin{equation}
F_S(q_0,\epsilon_+,\epsilon_-) = {1\over 2}\left[F(q_0,\epsilon_+,\epsilon_-)+F(q_0,\epsilon_-,\epsilon_+)\right]
\end{equation}
satisfies the symmetry when exchanging the quark energies $\epsilon_-$ and $\epsilon_+$. 

The exchange symmetry of the gluon self-energy helps us a lot for its divergence analysis. The low limit of the integration in (\ref{int}) is $\epsilon_- = \epsilon_+ =m_f$ and the up limit is $\epsilon_-=\epsilon_+=\infty$. Considering that $(\epsilon_--\epsilon_+)^2-{\bf q}^2$ in $\overline\Pi_{--}$ and $H^{\mu\nu}(q,\epsilon_+,\epsilon_-)$ in $\overline\Pi_{--}^{\mu\nu}$ are finite at the low limit and only finite polynomials at the up limit, the problem of divergence is controlled by the factor
\begin{equation}
\epsilon_-\epsilon_+F_S \sim {f_F(\epsilon_+\pm\mu_f)-f_F(\epsilon_-\pm\mu_f)\over q_0+\epsilon_+-\epsilon_-}
\end{equation}
which is finite at the low limit for any $q_0$ and goes to zero exponentially at the up limit. Therefore, there are no infrared (for massless quarks) and ultraviolet divergences for the self-energy $\Pi^{\mu\nu}_{--}$.         

\subsection{$\Pi_Q^{\mu\nu,ab}(q)$}
We now consider the contribution of the first loop in Figure \ref{fig1} constructed by two positive-energy quarks to the gluon self-energy. Similar to (\ref{cs}), we first separate the trivial color part from the spin and momentum dependent part, then take variable substitution $k\to -k$ and $\tilde k \to -\tilde k$. Considering the relations $\epsilon_{-k+q/2}=\epsilon_{k-q/2}=\epsilon_-, \epsilon_{-k-q/2}=\epsilon_{k+q/2}=\epsilon_+$ for the quark energies and $\Lambda_+(-\tilde k,m_f)=\Lambda_-(\tilde k,-m_f)$ for the projectors, and taking into account the property that the trace (\ref{trace1}) in spin space is symmetric under the exchange of $\mu$ and $\nu$ and the replacement of $m_f$ by $-m_f$, we have 
\begin{eqnarray}
&& \text{Tr}\left[\Lambda_+(-\tilde k+q/2)\gamma^0\gamma^\nu\Lambda_+(-\tilde k-q/2)\gamma^0\gamma^\mu\right]\nonumber\\
&=&\text{Tr}\left[\Lambda_-(\tilde k_-)\gamma^0\gamma^\mu\Lambda_-(\tilde k_+)\gamma^0\gamma^\nu\right],
\end{eqnarray}   
which leads to the result that the two loops constructed by positive- or negative-energy quarks have exactly the same contribution to the gluon self-energy,
\begin{equation}
\Pi_{++}^{\mu\nu,ab}(q) = \Pi_{--}^{\mu\nu,ab}(q).
\end{equation}

The third loop in Figure \ref{fig1} contains vacuum divergence, arising from the mixing between the positive- and negative-energy quarks. Excluding the same color factor, its contribution to the gluon self-energy is represented as
\begin{equation}
\Pi_{+-}^{\mu\nu} =  {g^2T\over 4V}\sum_{k,s=\pm}{\text{Tr}\left[\Lambda_+(\tilde k_+)\gamma^0\gamma^\mu\Lambda_-(\tilde k_-)\gamma^0\gamma^\nu\right]\over \left(k_+^0+s\mu_f-\epsilon_+\right)\left(k_-^0+s\mu_f+\epsilon_-\right)}.
\end{equation}
We take the trace in spin space
\begin{eqnarray}
\label{trace2}
&& \text{Tr}\left[\Lambda_+(\tilde k_+)\gamma^0\gamma^\mu\Lambda_-(\tilde k_-)\gamma^0\gamma^\nu\right]\nonumber\\
&=& {-1\over \epsilon_+\epsilon_-}\left[g^{\mu\nu}\left(m_f^2-\tilde k_+^\sigma \bar k^-_\sigma\right)+\tilde k_+^\mu \bar k_-^\nu+\tilde k_+^\nu \bar k_-^\mu\right]
\end{eqnarray}
and the summation over quark frequencies
\begin{eqnarray}
\label{j}
&& \sum_{k_0,s=\pm}{1\over\left(k_+^0+s\mu_f-\epsilon_+\right)\left(k_-^0+s\mu_f+\epsilon_-\right)}\\
&=& {\sum_{s=\pm}\left[f_F(\epsilon_++s\mu_f)+f_F(\epsilon_-+s\mu_f)\right]-2\over T\left(q_0+\epsilon_++\epsilon_-\right)},\nonumber\\
&\equiv& {\epsilon_-\epsilon_+\over 2T}\left[J(q_0,\epsilon_+, \epsilon_-)-{4\over \epsilon_-\epsilon_+\left(q_0+\epsilon_++\epsilon_-\right)}\right],\nonumber
\end{eqnarray}
$\Pi^{\mu\nu}_{+-}(q)$ can again be separated into two parts,
\begin{equation}
\Pi^{\mu\nu}_{+-} =g^{\mu\nu}\overline\Pi_{+-}+ \overline\Pi_{+-}^{\mu\nu}.
\end{equation}
Integrating out the azimuth angle $\phi$, performing variable substitution from $(|{\bf k}|,\cos\theta)$ to $(\epsilon_-,\epsilon_+)$, and considering the exchange symmetry between $\epsilon_-$ and $\epsilon_+$, the scalar and tensor functions $\overline\Pi_{+-}$ and $\overline\Pi_{+-}^{\mu\nu}$ are written as 
\begin{eqnarray}
\overline\Pi_{+-} &=& {g^2\over 32\pi^2}\int_R d\epsilon_-d\epsilon_+{\epsilon_-\epsilon_+\over |{\bf q}|}{{\bf q}^2-(\epsilon_-+\epsilon_+)^2\over 2}\nonumber\\
&&\times J_S(q_0,\epsilon_+,\epsilon_-),\nonumber\\
\overline\Pi_{+-}^{\mu\nu} &=& {g^2\over 32\pi^2}\int_R d\epsilon_-d\epsilon_+{\epsilon_-\epsilon_+\over |{\bf q}|}I^{\mu\nu}(q,\epsilon_+,\epsilon_-)\nonumber\\
&&\times J_S(q_0,\epsilon_+,\epsilon_-),\nonumber\\
I^{00} &=& H^{00},\nonumber\\
I^{0i} &=& I^{i0} ={{\bf q}^2-(\epsilon_+-\epsilon_-)^2\over 2|{\bf q}|}q_0\hat q^i,\nonumber\\
I^{ij} &=& -H^{ij},
\end{eqnarray}
where the function $J_S$ is defined as 
\begin{eqnarray}
J_S(q_0,\epsilon_+,\epsilon_-)={1\over 2}\left[J(q_0,\epsilon_+,\epsilon_-)+J(-q_0,\epsilon_+,\epsilon_-)\right]
\end{eqnarray}
with the function $J(q_0, \epsilon_+, \epsilon_-)$ defined in (\ref{j}). Like the function $F$, $J$ contains only the Fermi-Dirac distribution too, and therefore the integral is convergent at any temperature. However, we have here neglected the second term in the square bracket of (\ref{j}) which is temperature independent but leads to a divergence of $\overline\Pi_{+-}$ and $\overline\Pi_{+-}^{\mu\nu}$ when doing energy integral. Since the second term is medium independent, it behaves like the zero point energy of the harmonic oscillator in quantum mechanics and can then be removed safely when discussing thermodynamic properties of the system relative to vacuum. 
   
It can be checked that the total quark contribution to the gluon self-energy at one loop level, 
\begin{eqnarray}
\Pi_Q^{\mu\nu,ab} (q) &=& -{1\over T^2}{\delta_{ab}\over 2}\Pi_Q^{\mu\nu}(q),\nonumber\\
\Pi_Q^{\mu\nu}(q) &=& 2\Pi_{--}^{\mu\nu}(q)+2\Pi_{+-}^{\mu\nu}(q)\equiv \overline\Pi_Q^{\mu\nu}(q)
\end{eqnarray}
satisfies the Ward-identity~\cite{Xu:1991wp} at any temperature $T$, chemical potential $\mu_f$ and quark mass $m_f$,
\begin{equation}
q_\mu\overline\Pi_Q^{\mu\nu}(q)=0.
\end{equation}
 
The total quark loop $\overline\Pi_Q^{\mu\nu}$ can be separated into the longitudinal and transverse parts by using the tensor projectors $P_L^{\mu\nu}$ and $P_T^{\mu\nu}$~\cite{TFT_Bellac1996},  
\begin{equation}
\overline\Pi_Q^{\mu\nu}(q) = \overline\Pi_Q^T(q) P_T^{\mu\nu} + \overline\Pi_Q^L(q) P_L^{\mu\nu}.
\end{equation}
Using the relations $\overline\Pi_Q^{00}=(1-q_0^2/q^2)\overline\Pi_Q^L$ and $\overline\Pi_{Q\mu}^\mu=2\overline\Pi_Q^T+\overline\Pi_Q^L$, the transverse and longitudinal self-energies $\overline\Pi_Q^T(q)$ and $\overline\Pi_Q^L(q)$ are expressed as   
\begin{eqnarray}
\overline\Pi_Q^T &=& {g^2\over 4\pi^2}{1\over |{\bf q}|}\int_R d\epsilon_- d\epsilon_+\sum_{s=\pm} s{\epsilon_-+s\epsilon_+\over q_0^2-(\epsilon_-+s \epsilon_+)^2}\nonumber\\
&&\times\left\{m_f^2+{1\over 4{\bf q}^2}\prod_{s'=\pm}[(\epsilon_-+ss'\epsilon_+)^2-s'{\bf q}^2]\right\}\nonumber\\
&&\times f_F^s(\epsilon_+,\epsilon_-),\nonumber\\
\overline\Pi_Q^L &=& -{g^2\over 8\pi^2}{q^2\over |{\bf q}|^3}\int_R d\epsilon_- d\epsilon_+\sum_{s=\pm} s(\epsilon_-+s\epsilon_+)\nonumber\\
&&\times{(\epsilon_--s\epsilon_+)^2-{\bf q}^2\over q_0^2-(\epsilon_-+s \epsilon_+)^2}f_F^s(\epsilon_+,\epsilon_-),\nonumber\\
f_F^s &=& {1\over 2}\sum_{s'=\pm}\left[f_F(\epsilon_-+s'\mu_f)+s f_F(\epsilon_++s'\mu_f)\right].
\end{eqnarray}

\subsection{Gluon loop and ghost loop}
We now calculate the contribution from gluon loop and ghost loop to gluon self-energy. After taking the renormalization process to remove the divergence in vacuum, it satisfies the Ward identity too and can be represented by the projectors $P_L^{\mu\nu}$ and $P_T^{\mu\nu}$,
\begin{equation}
\overline\Pi_G^{\mu\nu}(q) = \overline\Pi_G^T(q) P_T^{\mu\nu}+\overline\Pi_G^L(q) P_L^{\mu\nu}
\end{equation}
with the transverse and longitudinal self-energies, 
\begin{eqnarray}
\overline\Pi_G^T &=& {3g^2 \over 8\pi^2}{1\over |{\bf q}|^3}\int_{R'}d\epsilon_- d\epsilon_+\sum_{s=\pm}s(\epsilon_-+s\epsilon_+)\nonumber\\
&&\times {q^2\left((\epsilon_--s\epsilon_+)^2+{\bf q}^2\right)+2{\bf q}^2\left((\epsilon_-+s\epsilon_+)^2-{\bf q}^2\right)\over q_0^2-(\epsilon_-+s\epsilon_+)^2}\nonumber\\
&&\times f_B^s(\epsilon_+,\epsilon_-),\nonumber\\
\overline\Pi_G^L &=& -{3g^2\over 4\pi^2}{q^2 \over |{\bf q}|^3}\int_{R'}d\epsilon_- d\epsilon_+\sum_{s=\pm}s(\epsilon_-+s\epsilon_+)\nonumber\\
&&\times {(\epsilon_--s\epsilon_+)^2-2{\bf q}^2\over q_0^2-(\epsilon_-+s\epsilon_+)^2}f_B^s(\epsilon_+,\epsilon_-),\nonumber\\
f_B^s &=& {1\over 2}\left[f_B(\epsilon_-)+s f_B(\epsilon_+)\right],
\end{eqnarray}
where $f_B(x)=1/(e^{x/T}-1)$ is the Bose-Einstein distribution, and the integration region $R'$ is the massless limit of the region $R$.

With the total gluon self-energy,
\begin{equation}
\overline\Pi^{\mu\nu}(q) = \overline\Pi_Q^{\mu\nu}(q) + \overline\Pi_G^{\mu\nu}(q),
\end{equation}
the gluon propagator under covariant gauge condition at one loop level has the well know form~\cite{TFT_Bellac1996,Weldon:1996kb}
\begin{equation}
\label{gp}
\Delta^{\mu\nu} = {P_T^{\mu\nu}\over q^2+\overline\Pi^T(q)}+{P_L^{\mu\nu}\over q^2+\overline\Pi^L(q)}+{\xi\over q^2}E^{\mu\nu}
\end{equation}
with $E^{\mu\nu}=q^\mu q^\nu/q^2,\ \overline\Pi^T=\overline\Pi^T_Q+\overline\Pi_G^T,\ \overline\Pi^L=\overline\Pi^L_Q+\overline\Pi_G^L$ and the gauge dependent parameter $\xi$.

\section{Application}
\label{sec4} 
\subsection{Debye mass}
In HTL and HDL approaches, the ring diagram resummation technique in QED and QCD~\cite{Kalashnikov:1979cy,Toimela:1982hv} is employed to consider non-perturbative effects. With the resummed gluon propagator, the Debye screening mass $m_D$ is defined as~\cite{Shuryak:1980tp,Gross:1980br,Kapusta:2006pm} 
\begin{equation}
m_D^2 = -\overline\Pi_{00}(q_0=0,|{\bf q}|\rightarrow 0).
\end{equation}
In extremely hot and dense QCD for massless quarks ($T, \mu_f\gg m_f$), it is represented as~\cite{TFT_Bellac1996,Schneider:2003uz}, 
\begin{equation}
\label{Debye}
m_D^2 = g^2\left({N_c\over 3}+{N_f\over 6}\right)T^2+g^2\sum_f{\mu_f^2\over 2\pi^2}
\end{equation}
with the number of colors $N_c$ and number of flavors $N_f$. 

We now take our obtained total gluon self-energy $\Pi^{\mu\nu}(q)$ to calculate the Debye screening mass in general case at finite temperature and density. We consider first the contribution from the loop constructed by two negative-energy quarks to the Debye mass,
\begin{equation}
m_{--}^2 = -g^{00}\overline\Pi_{--}(q_0=0,|{\bf q}|\rightarrow 0)-\overline\Pi_{--}^{00}(q_0=0,|{\bf q}|\rightarrow 0).
\end{equation}
In massless case with $m_f=0$ the integration region $R$ for the quark energies $\epsilon_+$ and $\epsilon_-$ is reduced to $R'$. For a function $A(q,\epsilon_+,\epsilon_-)$ with exchange symmetry between $\epsilon_+$ and $\epsilon_-$, the integration can be written as
\begin{eqnarray}
&& \int_{R'}d\epsilon_- d\epsilon_+ A(q,\epsilon_+,\epsilon_-)\nonumber\\
&=& \int_0^\infty d\epsilon_-\int_0^\infty d\epsilon_+A(q,\epsilon_+,\epsilon_-)\nonumber\\
&&\times\Theta\left(\epsilon_-+\epsilon_+-|{\bf q}|\right)\Theta\left(|{\bf q}|-|\epsilon_--\epsilon_+|\right)\nonumber\\
&=& 2\int_0^\infty d\epsilon_-\int_0^{\epsilon_-}d\epsilon_+ A(q,\epsilon_+,\epsilon_-)\nonumber\\
&&\times\Theta\left(\epsilon_-+\epsilon_+-|{\bf q}|\right)\Theta\left(|{\bf q}|-(\epsilon_--\epsilon_+)\right).
\end{eqnarray}
Taking Taylor expansion for the two step functions around $|{\bf q}|=0$,
\begin{eqnarray}
&& \Theta\left(\epsilon_-+\epsilon_+-|{\bf q}|\right)\Theta\left(|{\bf q}|-(\epsilon_--\epsilon_+)\right)\\
&=& \Theta(\epsilon_+-\epsilon_-)\Theta(\epsilon_++\epsilon_-)\nonumber\\
&&+{1\over |{\bf q}|}\left[\Theta(\epsilon_-+\epsilon_+)\delta(\epsilon_+-\epsilon_-)-\Theta(\epsilon_+-\epsilon_-)\delta(\epsilon_++\epsilon_-)\right],\nonumber
\end{eqnarray}
and considering the restriction $\epsilon_-\ge \epsilon_+$ and $\epsilon_- +\epsilon_+\ge 0$, only the second term with $\delta(\epsilon_+-\epsilon_-)$ contributes. From the limits
\begin{eqnarray}
&& \lim_{\epsilon_+\rightarrow\epsilon_-}\left[{1\over 2}(\epsilon_--\epsilon_+)^2+2\epsilon_-\epsilon_+\right] = 2\epsilon_-^2,\\
&& \lim_{\epsilon_+\rightarrow\epsilon_-}F_S(q_0=0,\epsilon_+,\epsilon_-)\nonumber\\
&=& -{2\over T\epsilon_-^2}\sum_{s=\pm}f_F(\epsilon_-+s\mu_f)\left(1-f_F(\epsilon_-+s\mu_f)\right)\nonumber,
\end{eqnarray}
we finally derive the expression for the Debye mass,
\begin{eqnarray}
m_{--}^2 &=& -{g^2\over 32\pi^2}\int_0^\infty d\epsilon_-\int_0^{\epsilon_-}d\epsilon_+ \epsilon_-\epsilon_+\delta(\epsilon_+-\epsilon_-)\nonumber\\
&&\times\left[(\epsilon_--\epsilon_+)^2+4\epsilon_-\epsilon_+\right]F_S(q_0=0,\epsilon_+,\epsilon_-)\nonumber\\
&=& {g^2\over 2} \left({T^2\over 6}+{\mu_f^2\over 2 \pi ^2}\right).
\end{eqnarray}

Taking into account the positive-energy quark loop $\Pi_{++}^{\mu\nu}$ which contributes the same to the Debye mass, the mixed quark loop $\Pi_{+-}^{\mu\nu}$ which makes no contribution due to the fact that the trace (\ref{trace2}) in spin space disappears in the limit $|{\bf q}|\to 0$ at $\mu=\nu=0$, the gluon and ghost loops $\Pi_G^{\mu\nu}$ which contributes \begin{equation}
m_G^2=-\Pi_G^{00}(q_0=0,|{\bf q}|\to 0)={N_c\over 3}g^2T^2,
\end{equation}
and the summation over all quark flavors, we obtain 
\begin{equation}
\label{mass}
m_D^2 = m_Q^2+m_G^2 = \sum_f \left(2m_{--}^2\right)+m_G^2,
\end{equation}
which is exactly the same as shown in (\ref{Debye}).

In general case with nonzero quark mass, the contribution from the negative-energy quark loop to the gluon self-energy at $q_0=0$ becomes
\begin{eqnarray}
\overline\Pi_{--} &=& \mathcal{O}\left(|{\bf q}|/m_f\right),\\
\overline\Pi_{--}^{0i} &=& \mathcal{O}\left(\left(|{\bf q}|/m_f\right)^2\right)\hat q^i,\nonumber\\
\overline\Pi_{--}^{00} &=& -{g^2m_f^3\over 8\pi^2T}\int_1^\infty dx x\sqrt{x^2-1}\nonumber\\
&&\times \sum_{s=\pm}f_F(x_s)\left(1-2f_F(x_s)\right)+\mathcal{O}\left(|{\bf q}|/m_f\right),\nonumber\\
\overline\Pi_{--}^{ij} &=& -\delta^{ij}{g^2m_f^3\over 24\pi^2T}\int_1^\infty dx {\left(x^2-1\right)^{3/2}\over x}\nonumber\\
&&\times\sum_{s=\pm} f_F(x_s)\left(1-2f_F(x_s)\right)+\mathcal{O}\left(|{\bf q}|/m_f\right)\nonumber
\end{eqnarray}
with $x_s=m_f x+s\mu_f$. Considering the similar contribution from the other quark loops and gluon and ghost loops, we derive the total Debye mass in the limit $|{\bf q}|=0$,
\begin{eqnarray}
m_D^2 &=& m_G^2+\sum_f{g^2m_f^3\over 4\pi^2T}\int_1^\infty dx x\sqrt{x^2-1}\nonumber\\
&&\times\sum_{s=\pm}f_F(x_s)\left(1-2f_F(x_s)\right),
\end{eqnarray}
it is reduced to the familiar result (\ref{mass}) for massless quarks.  

\subsection{High density limit}
Quark matter at low temperature and high baryon density can be realized in compact stars and nuclear collisions at intermediate energy. In the limit of zero temperature, the Bose-Einstein distribution $f_B$ disappears and the Fermi-Dirac distribution $f_F$ becomes a step function of the baryon chemical potential $\mu_f$. Therefore, the gluon self-energy comes only from the quark loops. For massless quarks, with the symmetric quark distributions
\begin{equation}
f^\pm_F(\epsilon_+,\epsilon_-) = \frac{1}{2} \left[\Theta\left(\mu_f-\epsilon_-\right)\pm \Theta\left(\mu_f-\epsilon_+\right)\right],
\end{equation}
and the continuation from Matsubara frequency summation to the Wick rotation integral $T\sum_{q_0}\to\int_{-\infty}^{+\infty}d\omega_q/(2\pi)$ $(q_0=-i\omega_q)$, the integration over the quark momentum ${\bf k}$ can be done easily, and the total transverse and longitudinal parts of the gluon self-energy can be explicitly expressed as
\begin{eqnarray}
\overline\Pi^T &=& -\frac{g^2 \mu_f^2}{12 \pi ^2}\left(\frac{2 q_0^2}{{\bf q}^2}+1\right)\nonumber\\
&& +\frac{g^2}{384 \pi ^2 |{\bf q}|^3}\sum_{n,s=\pm}{\cal F}_T(nq_0,|{\bf q}|,s\mu_f),\nonumber\\
\overline\Pi^L &=& \frac{g^2 \mu_f^2}{3 \pi ^2}\left(\frac{q_0^2}{{\bf q}^2}-1\right)\nonumber\\
&&-\frac{g^2}{192 \pi ^2 |{\bf q}|^3}\sum_{n,s=\pm}{\cal F}_L(nq_0,|{\bf q}|,s\mu_f)
\end{eqnarray}
with the functions ${\cal F}_T(nq_0,|{\bf q}|,s\mu_f)$ and ${\cal F}_L(nq_0,|{\bf q}|,s\mu_f)$ defined as 
\begin{eqnarray}
{\cal F}_T &=& \left(q_0^2+nq_0 |{\bf q}|+4 {\bf q}^2+4s \mu_f q_0+2 s\mu_f |{\bf q}|+4 \mu_f^2\right)\nonumber\\
&&\times\left(q_0^2-{\bf q}^2\right)\left(nq_0-|{\bf q}|+2s \mu_f\right)\nonumber\\
&&\times\ln {(nq_0-|{\bf q}|+2s\mu_f)^2\over (nq_0-|{\bf q}|)^2},\\
{\cal F}_L &=& \left(nq_0+2|{\bf q}|+2s\mu_f\right)\left(q_0^2-{\bf q}^2\right)\nonumber\\
&&\times\left(nq_0-|{\bf q}|+2s \mu_f\right)^2\ln {(nq_0-|{\bf q}|+2s\mu_f)^2\over (nq_0-|{\bf q}|)^2}.\nonumber
\end{eqnarray}

Under the condition of $q_0,|{\bf q}| \ll \mu_f$ at extremely high baryon density, which is usually considered in HDL approach, the above transverse and longitudinal self-energies are reduced to the well-known result~\cite{Rajagopal:2000wf,TFT_Bellac1996},
\begin{eqnarray}
\Pi^T &=& -{m_E^2\over 2}{q_0\over|{\bf q}|}\left[\left(1-{q_0^2\over{\bf q}^2}\right)H\left({q_0\over |{\bf q}|}\right)+{q_0\over|{\bf q}|}\right],\nonumber\\
\Pi^L &=& -m_E^2\left(1-{q_0^2\over{\bf q}^2}\right)\left[1-{q_0\over|{\bf q}|}H\left({q_0\over|{\bf q}|}\right)\right]
\end{eqnarray}
with the effective gluon mass $m_E$ and function $H(x)$  
\begin{eqnarray}
&& m_E^2 = {g^2\mu_f^2\over 2\pi^2},\nonumber\\
&& H(x)={1\over 2}\ln\left({1+x\over 1-x}\right).
\end{eqnarray}

One can also check the gluon screening mass and dynamical mass in quark matter at high density by calculating the following limits of the gluon self-energies,
\begin{eqnarray}
&& -\overline\Pi^T(q_0=0,|{\bf q}|\to 0) =0,\nonumber\\
&& -\overline\Pi^L(q_0=0,|{\bf q}|\to 0) = {g^2\mu_f^2\over 2\pi^2},\nonumber\\
&& -\overline\Pi^T(q_0\to 0,|{\bf q}|=0) = {g^2\mu_f^2\over 6\pi^2},\nonumber\\
&& -\overline\Pi^L(q_0\to 0,|{\bf q}|=0) = {g^2\mu_f^2\over6\pi^2}.
\end{eqnarray}

\subsection{Thermodynamic potential}
we now close the resummed gluon propagator to see the gluon contribution to the thermodynamic potential which controls the global behavior of the hot and dense quark-gluon plasma. The gluon contribution with ring diagrams can be written as~\cite{Smedback:2011xs} 
\begin{equation}
\Omega=\Omega_0 +\Omega_1 +\Omega_{ring},
\end{equation}
where $\Omega_0$ is the free gluon contribution. To use the standard definition of $\Omega_{ring}$, we have separated $\Omega_1$ with only one self-energy on the ring from $\Omega_{ring}$. Taking into account the separation of a general gluon propagator into a transverse and a longitudinal parts, see (\ref{gp}), $\Omega_{ring}$ can be divided into two parts~\cite{Andersen:1999sf} 
\begin{equation}
\Omega_{ring} = -N_gR_T-{N_g\over2}R_L.
\end{equation}

A calculation with gluon Matsubara frequency summation and nonzero quark mass using HTL/HDL approximation can be seen in Refs. \cite{Arnold:1994ps,Cvetic:2002ju,Andersen:1999va}. Here we calculate $R_L$ and $R_T$ with the derived gluon self-energy without any restriction to the medium temperature and quark chemical potential. To simplify the calculation we consider massless limit. After integrating out the azimuth angles $\phi$ and $\theta$ of the gluon momentum ${\bf q}$, the temperature $T$, chemical potential $\mu_f$ and coupling constant $g$ dependence of the transverse and longitudinal parts $R_T(T,\mu_f,g)$ and $R_L(T,\mu_f,g)$ can be explicitly expressed as 
\begin{eqnarray}
R_T &=&\int {d\omega_q\over2\pi} \int d|{\bf q}|{4\pi{\bf q}^2\over(2\pi)^3}\nonumber\\
&&\times\left[\ln\left(1-{\Pi^T(-i\omega_q,|{\bf q}|)\over \omega_q^2+{\bf q}^2}\right)+{\Pi^T(-i\omega_q,|{\bf q}|)\over \omega_q^2+{\bf q}^2}\right],\nonumber\\
R_L &=&\int {d\omega_q\over2\pi} \int d|{\bf q}|{4\pi{\bf q}^2\over(2\pi)^3}\nonumber\\
&&\times\left[\ln\left(1-{\Pi^L(-i\omega_q,|{\bf q}|)\over \omega_q^2+{\bf q}^2}\right)+{\Pi^L(-i\omega_q,|{\bf q}|)\over \omega_q^2+{\bf q}^2}\right].\nonumber\\
\end{eqnarray}

\begin{figure}
	\centering
	\includegraphics[width=0.5\textwidth]{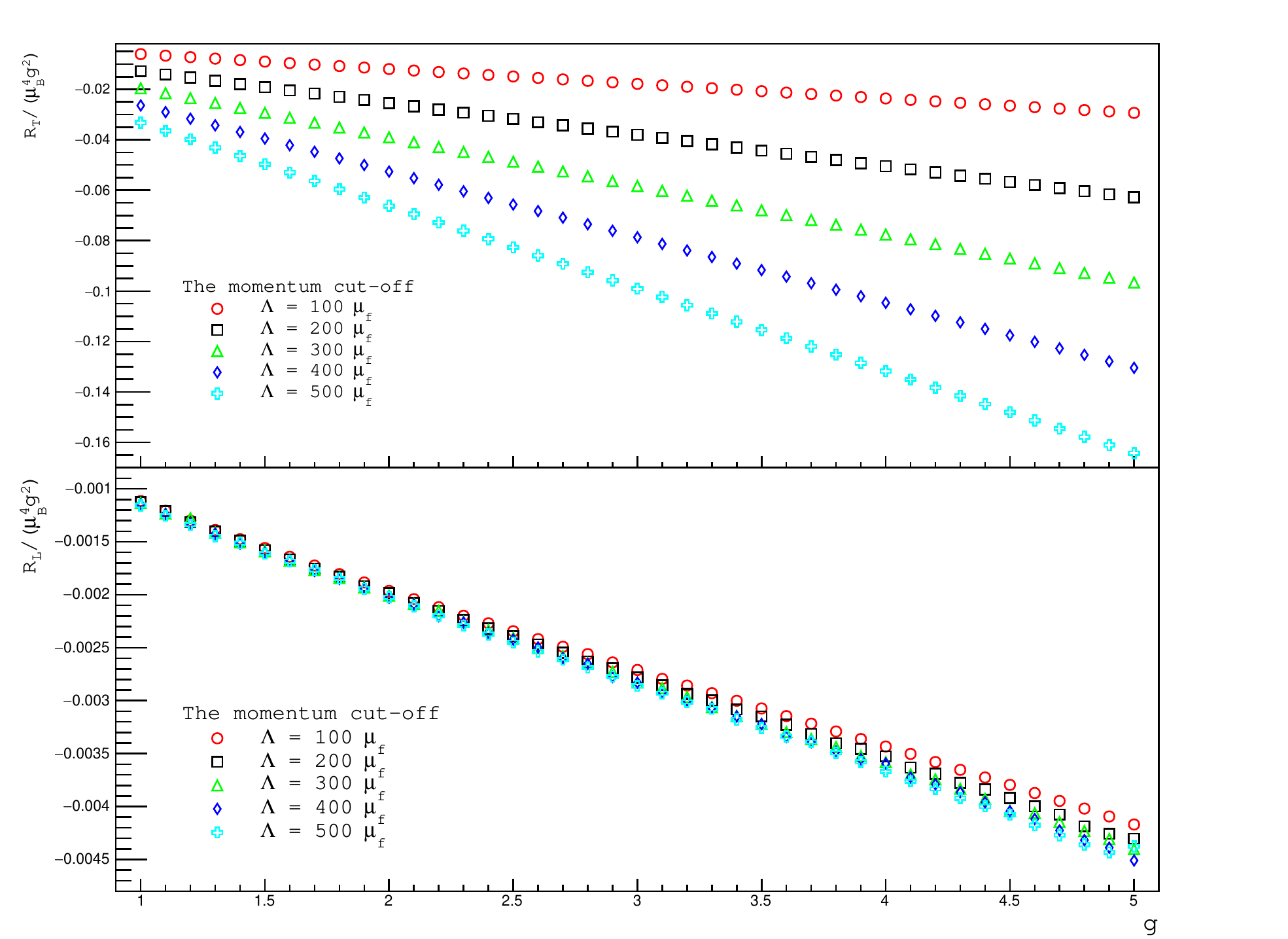}
	\caption{The scaled longitudinal and transverse thermodynamic potentials $R_L/g^2$ and $R_T/g^2$ as functions of coupling constant $g$ for a cold quark matter. The data points of different colors correspond to different gluon momentum cut-offs $\Lambda$.}
	\label{fig2}
\end{figure}

Considering $\Pi\sim g^2$, the Taylor expansion of the logrithm function leads to $R_T, R_L\sim g^4$ in weak coupling region. In the range $1<g<5$, the detailed calculation shows that $R_T/\mu_{\rm B}^4,R_L/\mu_{\rm B}^4\sim g^2+\lambda\ g^3$. In general case the coupling constant dependence of $R_L$ and $R_T$ for cold quark matter ($T=0$) at different gluon momentum cut-offs are shown in Figure \ref{fig2}. 

\section{Summary}
\label{sec5}
We focused in this paper on gluon self-energy at finite temperature, chemical potential and quark mass in the frame of QCD. With the method of quark energy-projectors, the divergence of the self-energy $\Pi^{\mu\nu}(q)$ appears only in the mixed loop constructed by the positive- and negative-energy modes and is medium independent. Therefore, the divergence can be removed through a renormalization procedure in vacuum. After a resummation of the gluon self-energy which is often used in literature, we calculated non-perturbatively the gluon Debye mass and thermodynamic potential. Since the Debye mass is irrelevant to the mixed quark loop, it is renormalization independent. In the limits of massless quarks and extremely high temperature and high density, our calculations come back to the well-known results from HTL and HDL. The quark energy-projector method used in this paper can be straightforwardly extended to treat the divergence of other QCD diagrams with more quark loops. 

{\bf Acknowledgement}: We thank Lianyi He for helpful discussions in the beginning of the work. The work is supported by Guangdong Major Project of Basic and Applied Basic Research No. 2020B0301030008 and the NSFC under grant Nos. 11890712 and 12075129.

\end{document}